\documentclass[twocolumn,aps,pra,10pt,showpacs,showkeys,superscriptaddress]{revtex4}

\usepackage{amsmath,amssymb}
\usepackage{graphicx}
\usepackage{dcolumn}
\usepackage{bm}
\usepackage{multirow}
\usepackage{float}

 \newcounter{abcd}
 
 \newcommand{\eqn}[1]{Eq.~(\ref{#1})}

 \newcommand{\stdp}{\strut\displaystyle} 
 \newcommand{\px}[1]{\stdp\frac{\partial}{\partial#1}}

\begin{document}

\title{A new time-frequency method to reveal quantum dynamics of atomic hydrogen in intense laser pulses: Synchrosqueezing Transform}


%

\author{Yae-lin Sheu}
\email{ispot.sheu@gmail.com}
\affiliation{Center for Quantum Science and Engineering, and Center for Advanced Study in Theoretical Sciences, Department of Physics, National Taiwan University, Taipei 10617, Taiwan}
\author{Liang-Yan Hsu}
\affiliation{Department of Chemistry, Princeton University, Princeton, New Jersey 08544, USA}
\author{Hau-tieng Wu}
\affiliation{Department of Mathematics, University of Toronto, Toronto ON M5S 2E4, Canada}
\author{Peng-Cheng Li}
\affiliation{Center for Quantum Science and Engineering, and Center for Advanced Study in Theoretical Sciences, Department of Physics, National Taiwan University, Taipei 10617, Taiwan}
\affiliation{College of Physics and Electronic Engineering, Northwest Normal University, Lanzhou 730070, China}
\author{Shih-I Chu}
\email{sichu@phys.ntu.edu.tw}
\affiliation{Center for Quantum Science and Engineering, and Center for Advanced Study in Theoretical Sciences, Department of Physics, National Taiwan University, Taipei 10617, Taiwan}
\affiliation{Department of Chemistry, University of Kansas, Lawrence, Kansas 66405, USA}

\begin{abstract}
This study introduces a new adaptive time-frequency (TF) analysis technique, synchrosqueezing transform (SST), to explore the dynamics of a laser-driven hydrogen atom at an {\it ab initio} level, upon which we have demonstrated its versatility as a new viable venue for further exploring quantum dynamics. For a signal composed of oscillatory components which can be characterized by instantaneous frequency, the SST enables rendering the decomposed signal based on the phase information inherited in the linear TF representation with mathematical support. Compared with the classical type TF methods, the SST clearly depicts several intrinsic quantum dynamical processes such as selection rules, AC Stark effects, and high harmonic generation. 
\end{abstract}


\maketitle

\noindent

Time-frequency (TF) analysis is a powerful tool for probing dynamics and has been extensively applied to macroscopic or classical dynamical systems, \textit{e.g.}, the cardiovascular system \cite{cardiovascular}, mitochondrial oscillations \cite{mitochondria}, and seismic data \cite{seismo}. Nevertheless, TF analysis on quantum systems has not yet received such attention. Recently, high-order harmonic generation (HHG), a fundamental quantum dynamical process, has attracted broad interest due to the advances in attosecond science and its potential application to tomographic imaging of molecular orbitals \cite{tomographic_imaging,molecular_orbital_tomography}, exploration of multichannel dynamics in strong-field ionization \cite{multichannel_dynamics}, etc.

To optimize the HHG processes, a detailed understanding of the corresponding dynamics and spectral structures is essential. 
Via the concept of instantaneous frequency (IF), the TF representation can reveal the emission time, intensity, phase, and interference \cite{ HH_imaging_few_cycle, Tong_Morlet, attosecond_physics_review,Time_profile_of_harmonic_generation} of the harmonics from the time-dependent induced dipole.
Up to date, HHG has been investigated by using the Gabor transform \cite{emission_times_in_HHG, Continuum_Wave_Packet_Interference, HHG_two_color_mutually_orthogonal}, the Morlet wavelet transform \cite{tomographic_imaging, Tong_Morlet}, and the Hilbert transform based on the notion of intrinsic mode functions (IMF) \cite{HHG_hilbert}. 
In this Letter, we first present a comprehensive study of TF methods on a quantum dynamical system, a hydrogen atom in an intense oscillating laser field, at the \textit{ab initio} level. We demonstrate the limitation of the Gabor transform and Morlet wavelet transform, as examples for the short time Fourier transform (STFT) \cite{Book_TF_analysis} and the continuous wavelet transform (CWT)  \cite{Ten_Lectures_on_Wavelets}, respectively, and the Wigner-Ville (WV) transform \cite{Book_TF_analysis}, as an example for the quadratic type TF method.
In order to circumvent the limitation of the above-mentioned methods, we introduce the synchrosqueezing transform (SST), which has successfully depicted chronotaxic systems \cite{Nonautonomous_Oscillators}, cardiovascular systems \cite{Wu_ECG_breathing_dynamics}, and breathing dynamics \cite{Wu_Ventilator_Weaning}.

The SST is proposed to address the intrinsic blurring in the linear type methods, such as the STFT \cite{Book_TF_analysis,Ten_Lectures_on_Wavelets} and CWT, and the accuracy and limitation of the SST technique are well supported by mathematical analysis \cite{Wu_thesis,Wu_sst_trend}. In a nutshell, based on the chosen linear TF representation, which can be STFT or CWT, denoted as $R(t,\omega)$, the SST sharpens the resolution of $R(t,\omega)$ by re-allocating its value at $(t,\omega)$ to a different point $(t,\omega')$ according to the reallocation rule determined from the phase information in $R(t,\omega)$. When the signal is composed of several oscillatory components with slowly time-varying amplitudes and frequencies that satisfy the constraints to be described, the SST enables decomposition of these oscillatory components. We mention that the properties of the SST oscillatory components, which share characteristics similar to those considered in the empirical mode decomposition, have been well studied theoretically \cite{Wu_thesis,Wu_sst_trend}. In particular, the notion of {\it instantaneous frequency} can be rigorously defined in the context of the SST.

The dynamics of hydrogen and the HHG processes can be understood via the time-dependent induced dipole in length form $d_L (t)$ \cite{Tong_gps}. In numerical simulations, we obtain the wave function $\psi({\bf r},t)$ at the position ${\bf r}$ and time $t$ by solving the time-dependent Schr\"odinger equation of the Hydrogen atom: $i\px{t}\psi({\bf r},t)=[H_0({\bf r},t)+V({\bf r},t)]\psi({\bf r},t)$, where $H_0 ({\bf r},t)$ is an unperturbed atomic Hamiltonian and $V({\bf r},t)=-zE_0 E(t)  \sin(\omega_0 t)$  is the interaction of the electron with the applied laser field with $z$ polarization. Note that $\omega_0$, $E_0$, and $E(t)$ are the fundamental frequency, amplitude, temporal profile of the laser pulse, respectively.  The simulation is carried out using a time-dependent generalized pseudospectral method \cite{Tong_gps}. 



In this study, we choose the modified STFT as the underlying method for the SST and give a brief introduction of the SST as follows. For a fixed window function $g$ in the Schwartz class $\mathcal{S}$, we denote the STFT of the signal $f(t)\in L^\infty (\mathbb{R})\cap C^1 (\mathbb{R})$ by
\begin{equation}
V_f(t,\eta)\equiv\int_{-\infty}^\infty f(x)g(x-t)e^{-i2\pi\eta(x-t)}\,dx.\label{eq.04}
\end{equation}
The decomposed components of the signal $f(t)$, regarded as the intrinsic mode type (IMT) functions, are defined as follows:\\

{\it Definition 1}\;: (IMT functions)

The space ${\cal B}_\varepsilon\subset L^\infty (\mathbb{R})\cap C^1 (\mathbb{R})$, where $0<\varepsilon\ll 1$, of the IMT functions consists of the functions $f_k:\mathbb{R}\rightarrow\mathbb{R}$ having the form
 \setcounter{abcd}{\arabic{equation}}
 \addtocounter{abcd}{1}
 \setcounter{equation}{0}
 \renewcommand{\theequation}{\arabic{abcd}\alph{equation}}
\begin{equation}
f_k (t)=A_k (t)\cos(2\pi\phi_k (t)),\label{eq.05a}
\end{equation}
such that $A_k (t)$ and $\phi_k (t)$ satisfy the following conditions:
\begin{equation}
\begin{array}{c}
A_k\in C^1(\mathbb{R}),\quad\phi_k\in C^2(\mathbb{R}),\quad A_k (t)>0,\\
\inf_{t\in\mathbb{R}}\phi'_k (t)>0,\quad\sup_{t\in\mathbb{R}}\phi'_k(t)<\infty\\
\left|A'_k (t)\right|\le\varepsilon\left|\phi'_k (t)\right|,\quad\left|\phi''_k (t)\right|\le\varepsilon\left|\phi'_k (t)\right|,\quad\forall\;t,\\
\mbox{and}\quad\sup_{t\in\mathbb{R}}\left|\phi''_k (t)\right|<\infty,
\end{array}\label{eq.05b}
\end{equation}
where the subscripts $'$ and $''$ denote the first- and the second-order derivatives with respect to t, respectively. Here $A_k (t)$ and $\phi'_k (t)$ are regarded as the amplitude modulation (AM) function and the instantaneous frequency (IF) function, respectively, of the IMT function $f_k(t)$ \cite{Wu_sst_trend}. Note that an IMT function can be viewed as a generalization of the harmonic function in that locally its amplitude and frequency are almost constant. Also, IMT function serves as a mathematical formula for the IMF considered in the EMD algorithm.\\  

{\it Definition 2}\;: (Superposition of the IMT functions)

The space ${\cal B}_{\varepsilon,d}\subset L^\infty(\mathbb{R})\cap C^1 (\mathbb{R})$, where $0<\epsilon\ll 1$ and $d>0$, of the superpositions of the IMT functions consists of the functions $f$ having the form

 \setcounter{equation}{\arabic{abcd}}
 \renewcommand{\theequation}{\arabic{equation}}
\begin{equation}
f(t)=\sum_{k=1}^K f_k (t),	\label{eq.06}
\end{equation}
for some finite $K>0$ and $f_k (t)=A_k (t)\cos(2\pi\phi_k (t))\in{\cal B}_\varepsilon$, such that the $\phi_k (t)$ satisfies $\phi'_k(t)-\phi'_{k-1} (t)>d$. (In another word, functions in ${\cal B}_{\varepsilon,d}$ are composed of several oscillatory components with slowly time-varying AM and IF, i.e., $\phi'_k (t)$, and the IF of any consecutive components are separated by at least $d$.) Note that ${\cal B}_\varepsilon$ and ${\cal B}_{\varepsilon, d}$ are not vector spaces.

According to {\it Definitions 1} and {\it 2}, we consider the reallocation rule derived from the phase information of the STFT as follows:\\

{\it Definition 3}\;: (Reallocation rule function)

Let $f(t)\in {\cal B}_{\varepsilon,d}$. Choose a window function $g\in \mathcal{S}$, such that $\mbox{supp}(\hat g)\subset\left[-\stdp\frac{d}{2},\stdp\frac{d}{2}\right]$, where $\hat g$ is the Fourier transform of $g$. The reallocation rule function $\omega_f (t,\eta)$ is defined as
\begin{equation}
\omega_f (t,\eta)=\left\{\begin{array}{ll}
 \stdp\frac{-i\partial_t V_f (t,\eta)}{2\pi V_f (t,\eta)} &\mbox{when}\quad V_f (t,\eta)\ne0\\
 \infty & \mbox{when}\quad V_f (t,\eta)=0.\end{array}\right.	\label{eq.07}
\end{equation}
As a special case in the reallocation method \cite{Wu_thesis,Book_TF_analysis}, the STFT synchrosqueezing method is given as:
\begin{equation}
S_f^{\alpha,\gamma} (t,\xi)\equiv\int_{A_{\gamma,f(t)}}V_f (t,\eta)\frac{1}{\alpha}
h\left(\frac{|\xi-\omega_f (t,\eta)|}{\alpha}\right)\,d\eta,\label{eq.08}
\end{equation}
where $\gamma,\alpha>0$, $h(t)=\frac{1}{\sqrt\pi} e^{-t^2}$ and $A_{\gamma,f(t)}\equiv\left\{\eta\in\mathbb{R}_+:\left|V_f (t,\eta)\right|>\gamma \right\}$. Note that the threshold $\gamma$ is to avoid numerical instability when computing \eqn{eq.07}. $\gamma$ also serves as a threshold when noise exists \cite{Wu_sst_trend}.  The main idea of the SST in \eqn{eq.08} regards that, at a certain time $t$, the values of $V_f (t,\eta)$ along the positive frequency axis $\eta$ are re-assigned to a new location $\xi$ by the delta-function like function $h(t)$ according to the reallocation rule function in \eqn{eq.07}. As a consequence, the SST can provide a substantially improved resolution in the TF representation. 

\begin{figure*}

	  \includegraphics[width=0.33\hsize]{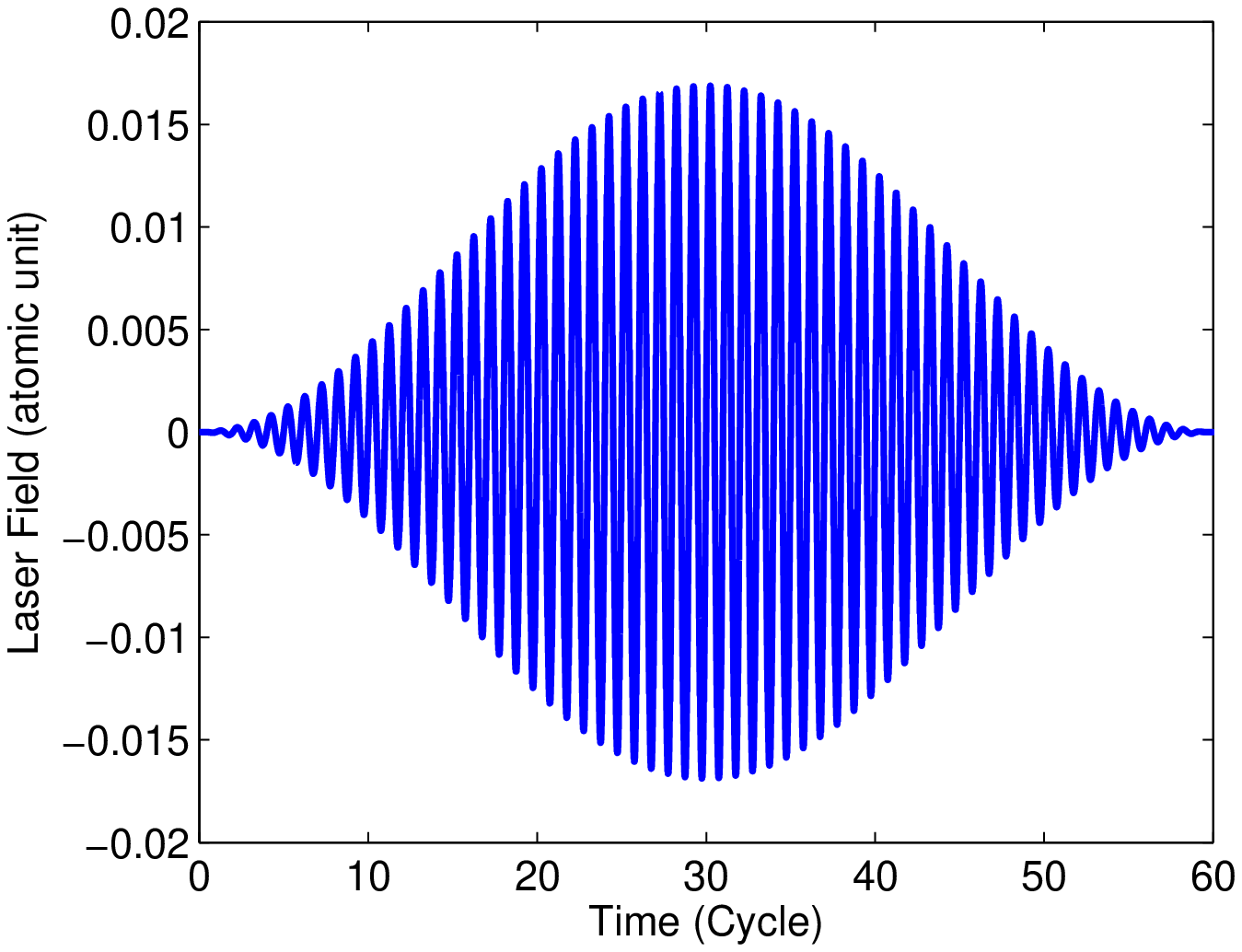}\hfill
		\includegraphics[width=0.33\hsize]{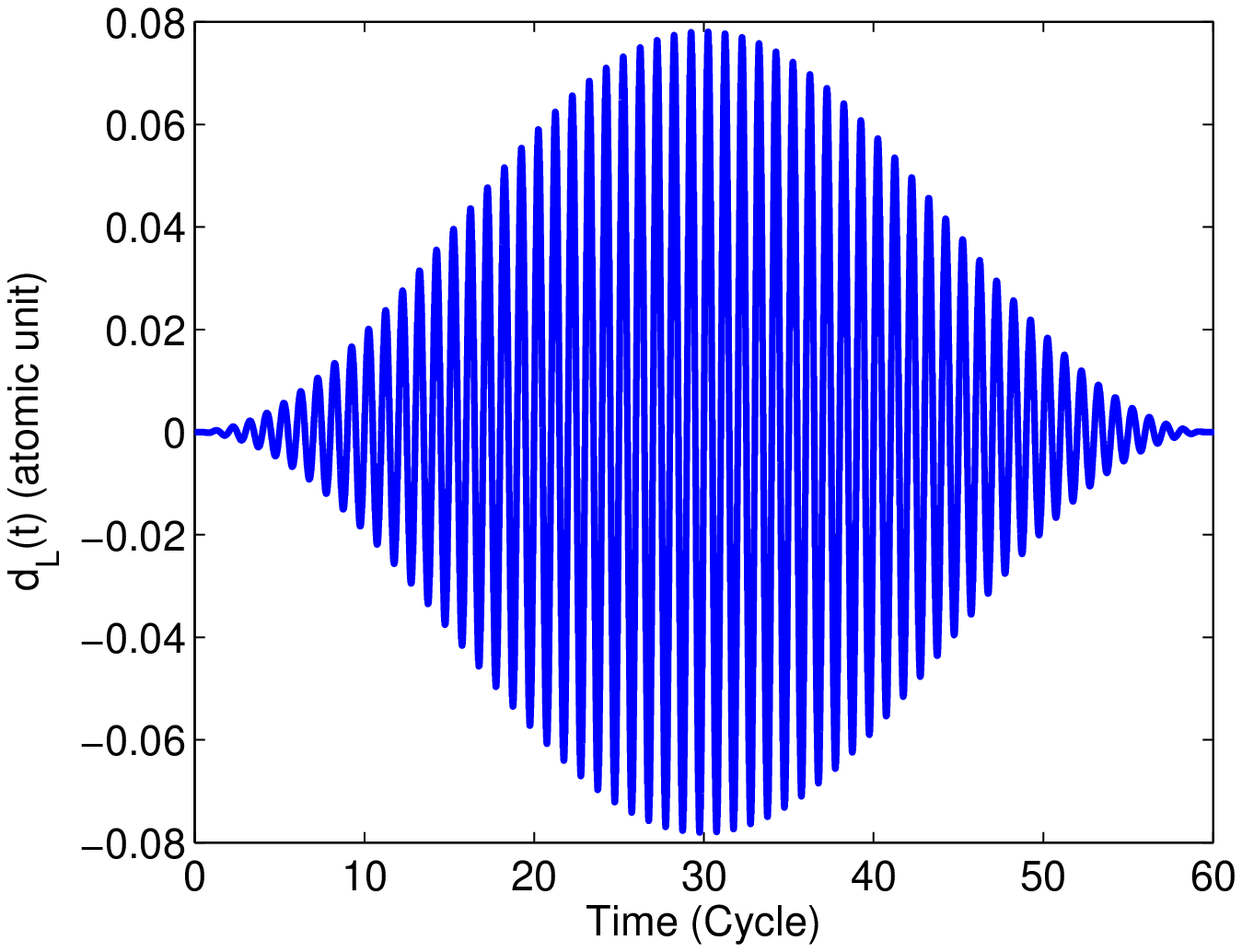}\hfill
    \includegraphics[width=0.33\hsize]{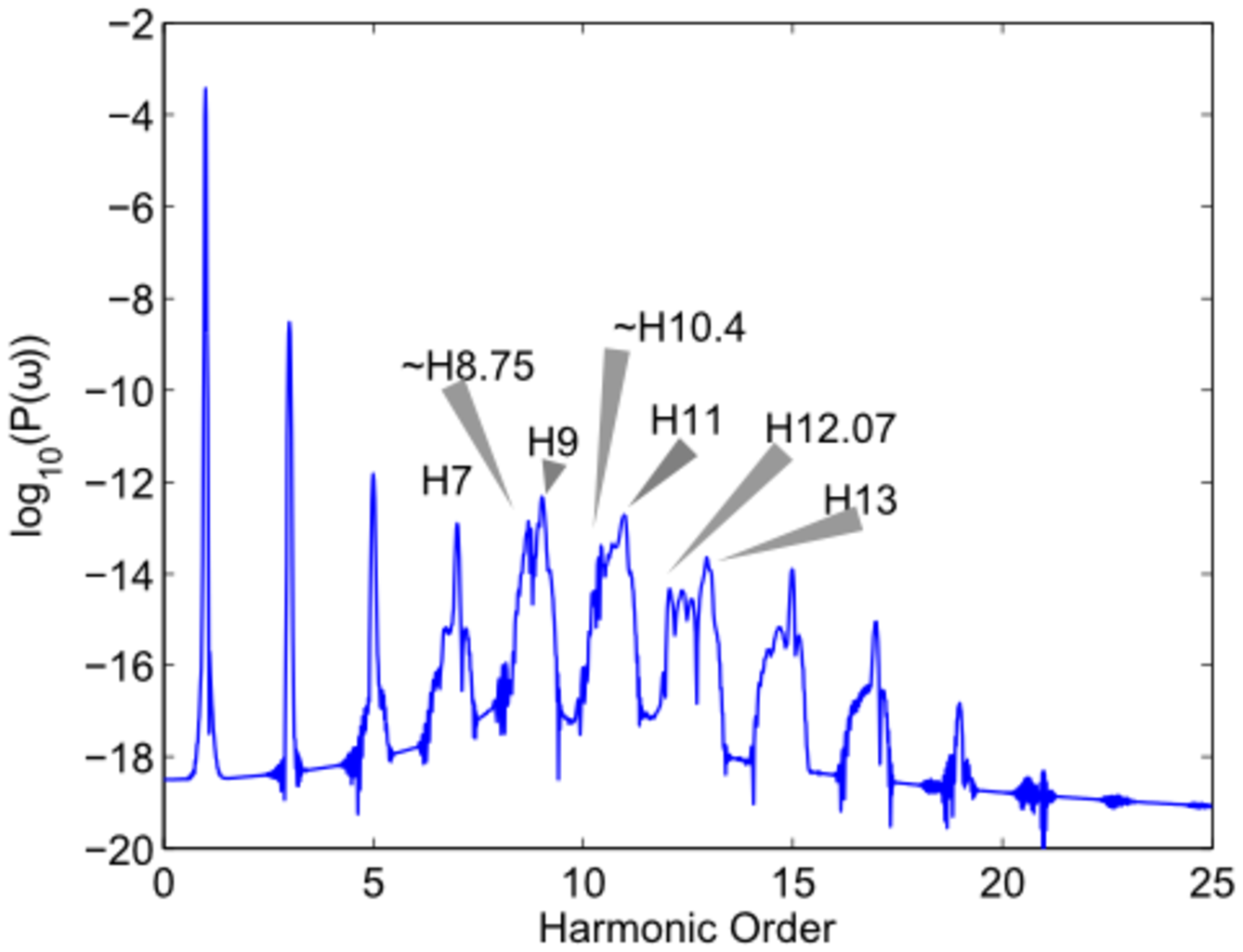}\\
\hspace*{0.2cm}(a)\hspace{5.6cm}(b)\hspace{5.4cm}(c)\\

	  \includegraphics[width=0.33\hsize]{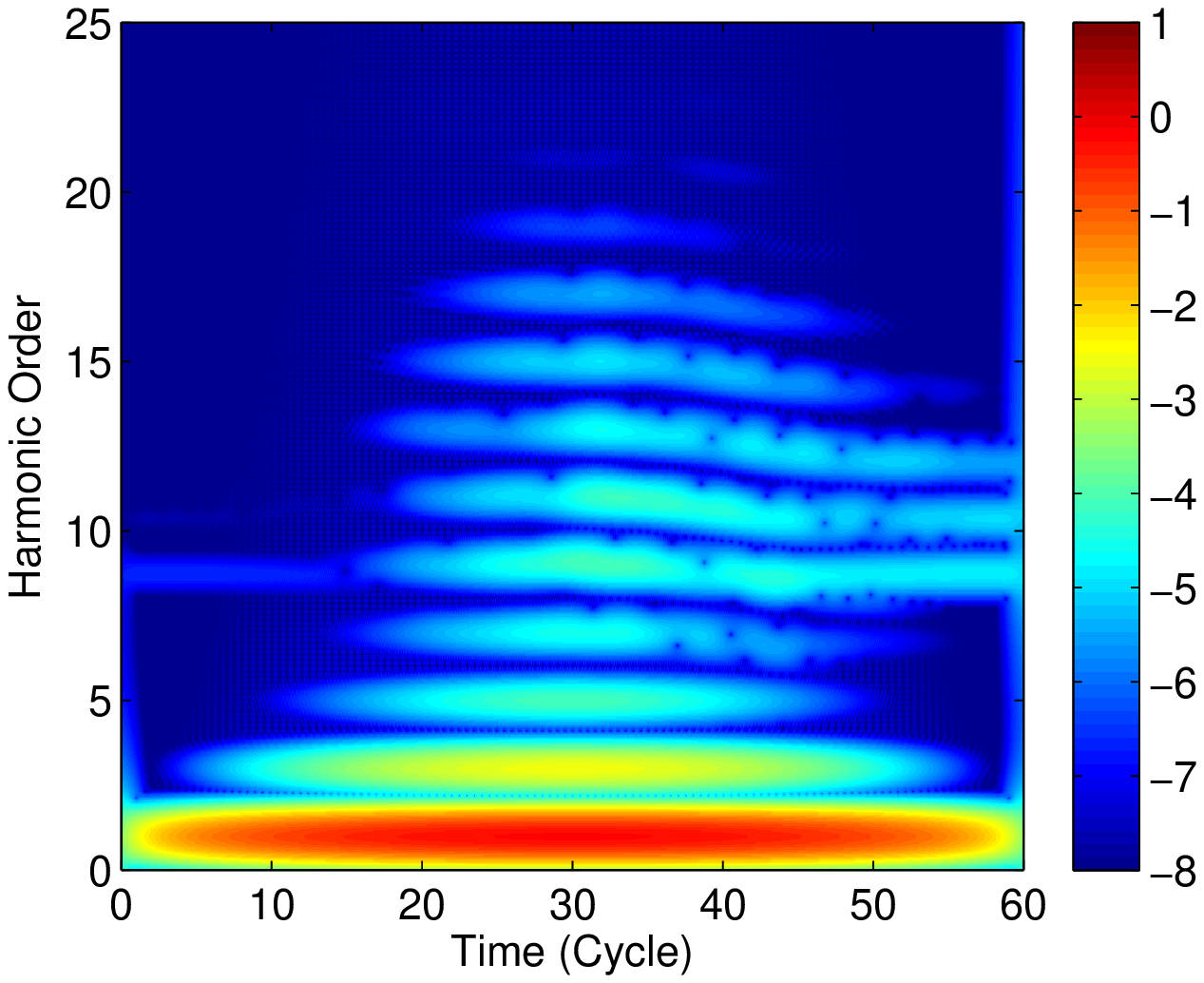}\hfill
		\includegraphics[width=0.33\hsize]{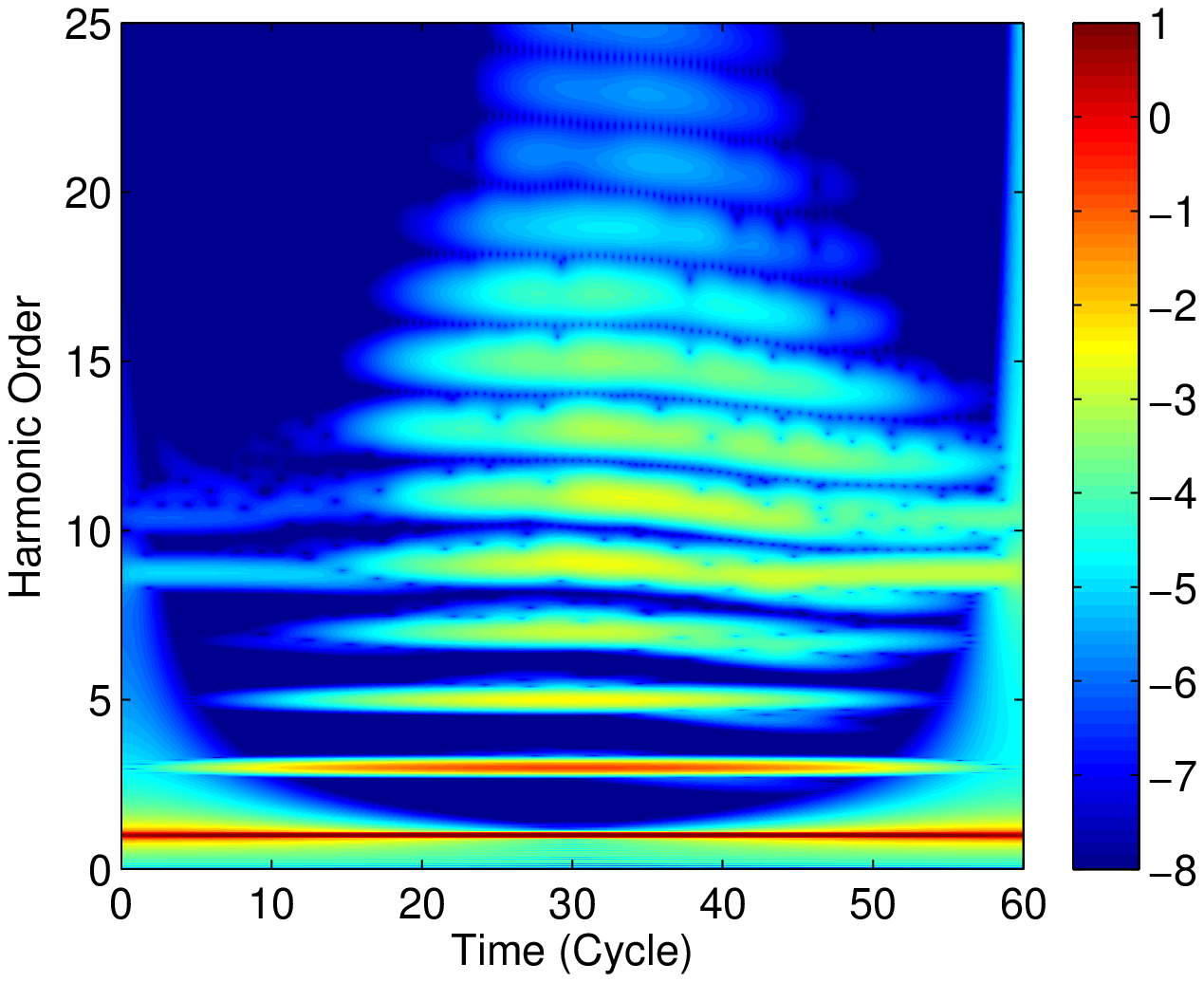}\hfill
		\includegraphics[width=0.33\hsize]{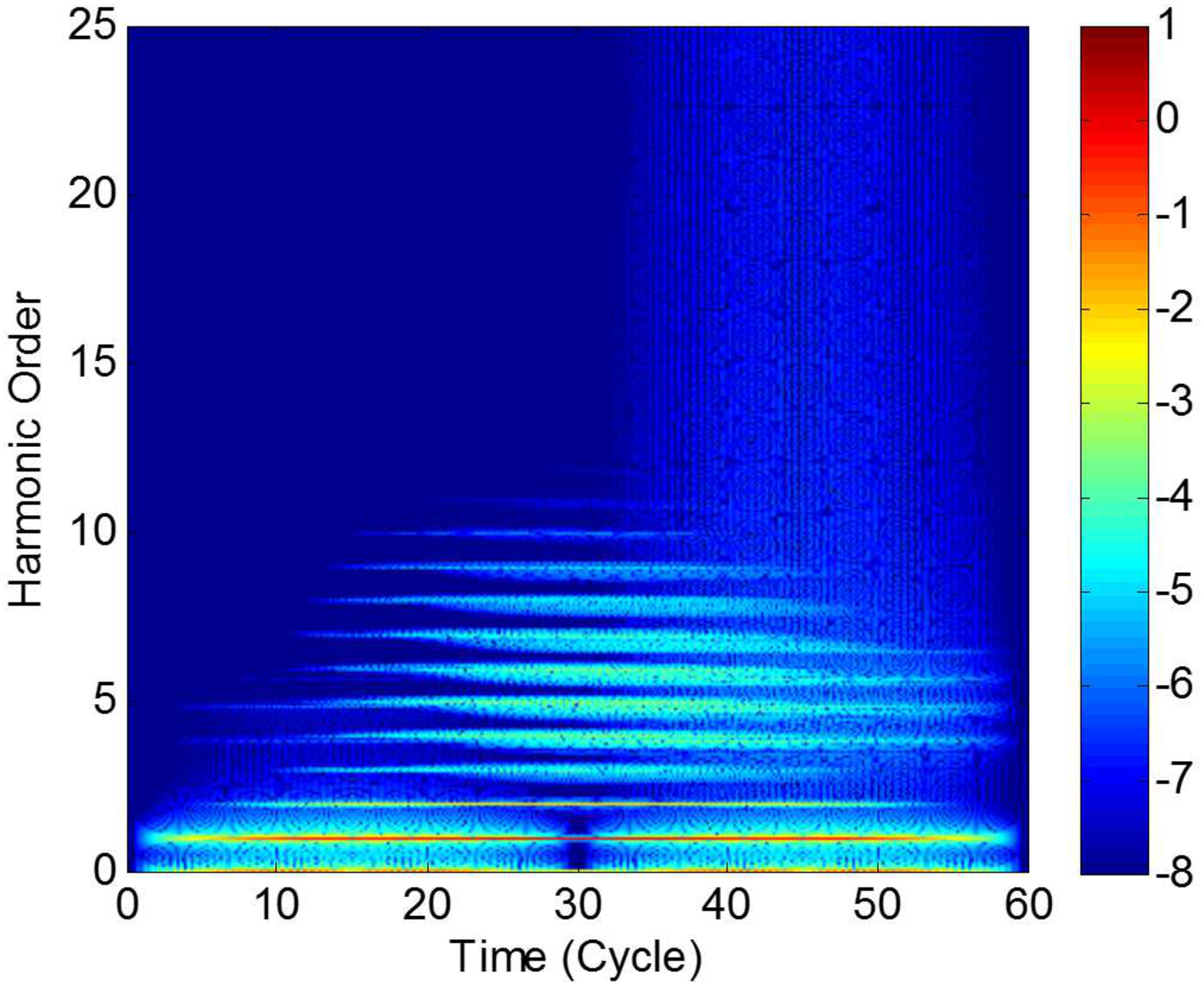}\\
\hspace*{0.2cm}(a)\hspace{5.6cm}(b)\hspace{5.4cm}(c)\\

	  \includegraphics[width=0.33\hsize]{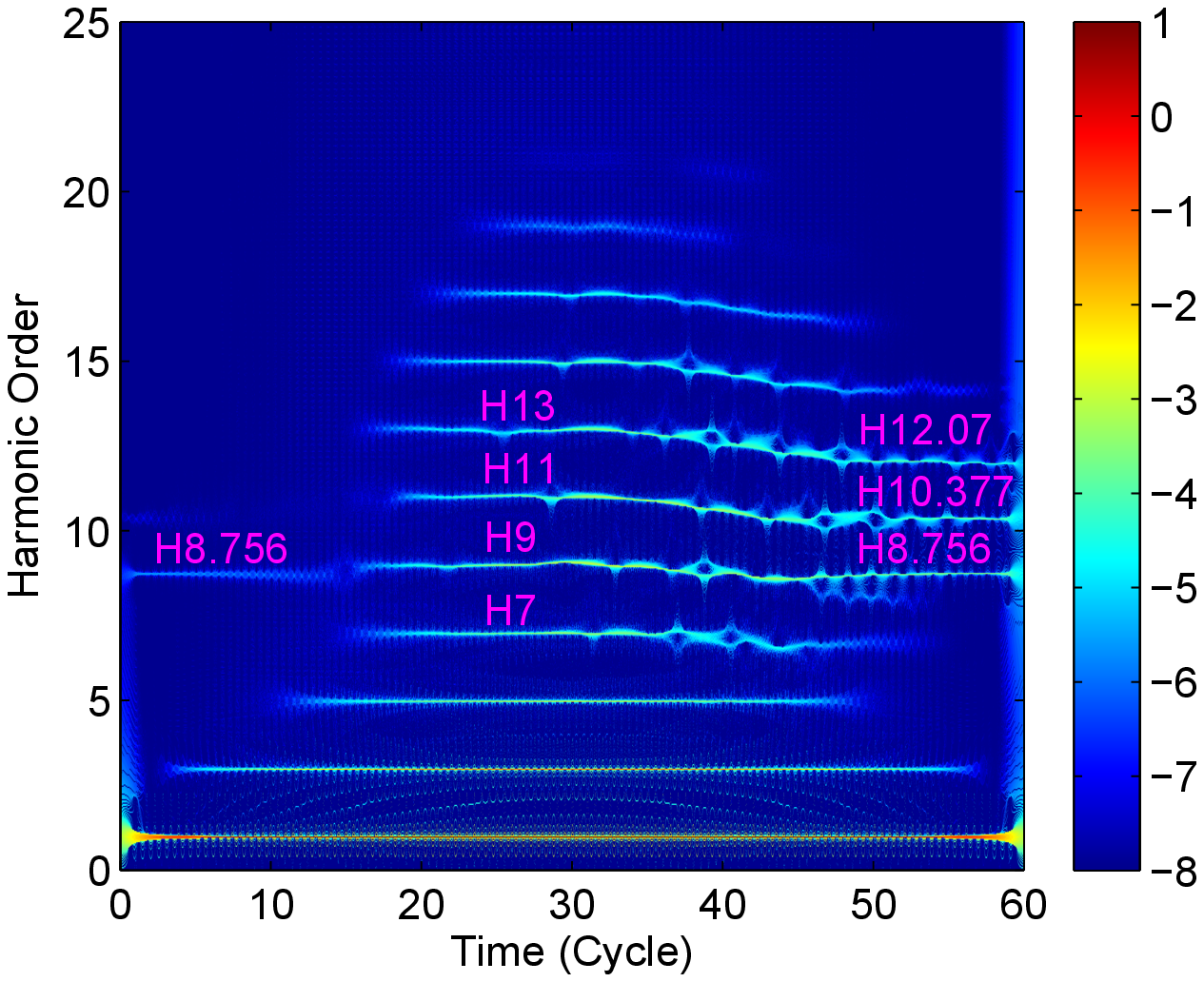}\hfill
		\includegraphics[width=0.27\hsize]{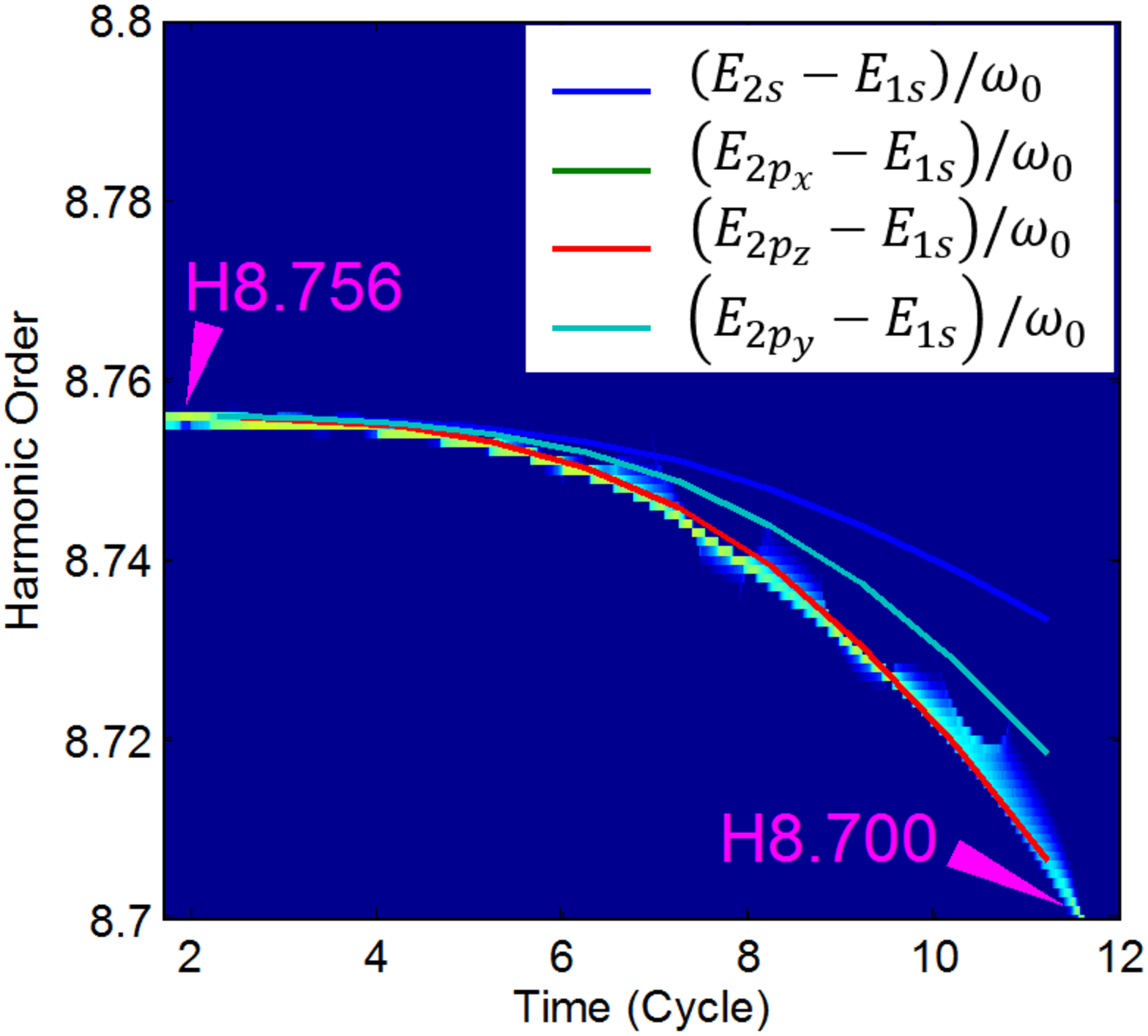}\hfill
		\includegraphics[width=0.37\hsize]{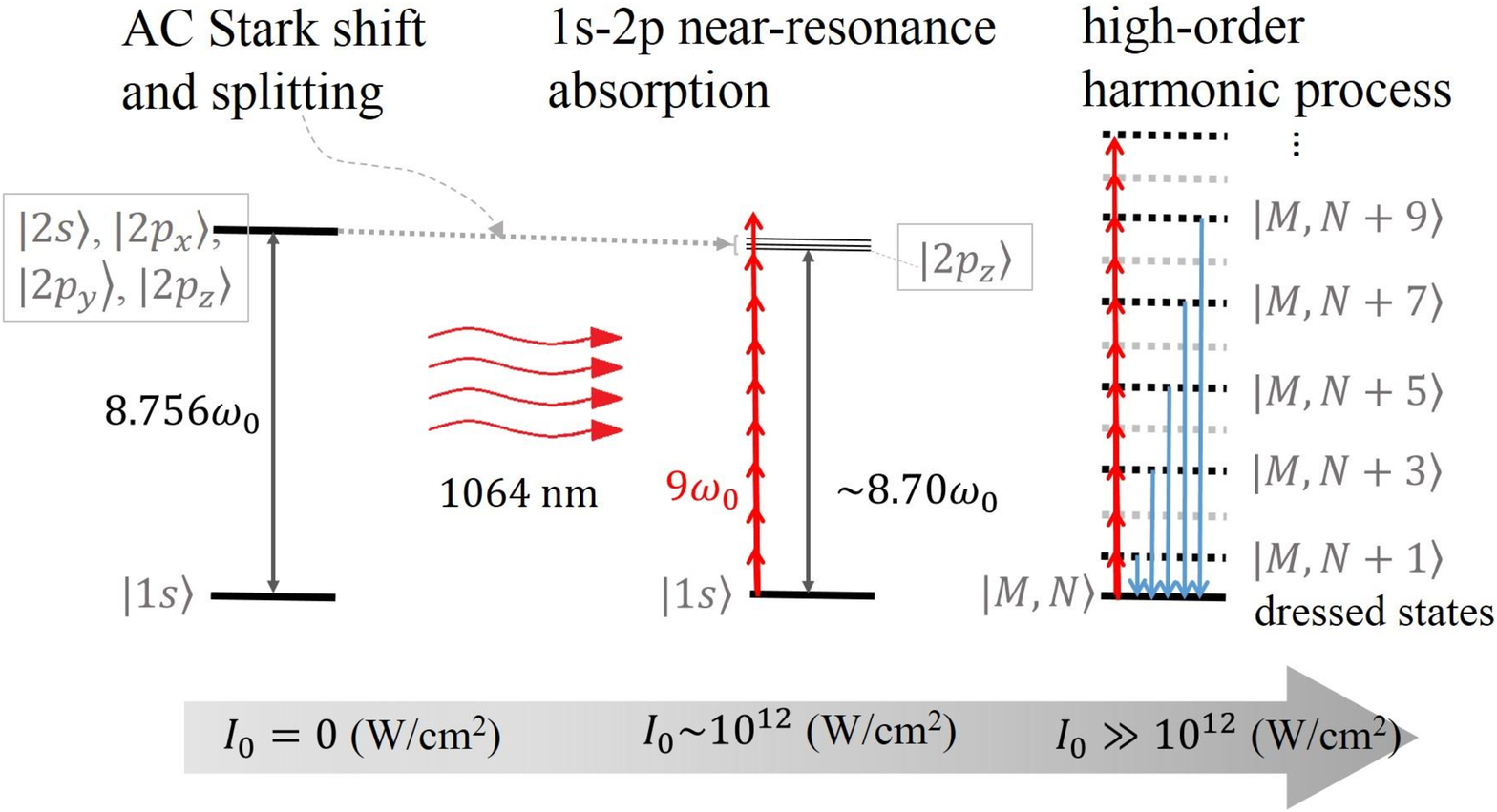}\\
\hspace*{0.2cm}(a)\hspace{5.6cm}(b)\hspace{5.4cm}(c)\\

		

 \caption{ (a) The laser field. (b) $d_L(t)$. (c) $P(\omega)$. TF representation by (d) Gabor, (e) Morlet, (f) WV transforms, and (g) SST. (h) The SST result and Floquet computation for 1s-2$\rm{p}_z$ transition. (i) The physical  mechanisms.
 }
  \label{Fig1}
\end{figure*}

In the simulation, we chose a laser field (Fig.~\ref{Fig1}(a)) $E_0E(t)\sin(\omega_0t)$ with a profile of $E(t)=\sin^2 (\pi t/(nT))$, where $n=60$ is the pulse length measured in optical cycles ($T=2\pi/\omega_0$), $\omega_0\approx0.0428278$ in atomic units (a.u.) corresponds to laser wavelength 1064 nm, and $E_0\approx0.0169$ in a.u. corresponds to the laser intensity of $I_0=10^{13}\;\mbox{W/cm}^2$. Figure \ref{Fig1}(b) presents the computed induced dipole in length form $d_L(t)$, and Fig.~\ref{Fig1}(c) shows the corresponding power spectrum obtained by $P(\omega)=\Big|\frac{1}{t_f-t_i}\int_{t_i}^{t_f}d_L(t)e^{-i\omega t}dt\Big|^2 $, where $t_i$ and $t_f$ is the initial and the final time of $d_L(t)$, respectively \cite{Tong_gps}. Note that the profiles of $d_L(t)$ and the applied laser field appear similar in the time domain, but they are different in the frequency domain. For example, while the power spectrum of the laser field has only one peak located at $\omega_0$, that of $d_L(t)$ reveals odd harmonics due to the parity symmetry \cite{book_attosecond}. However, the meaning of the substructures within the odd harmonics and their corresponding dynamics are unclear. 


To unveil the dynamics of $d_L(t)$, we apply several TF methods, namely, Gabor (Fig.~\ref{Fig1}(d)), Morlet (Fig.~\ref{Fig1}(e)), and WV (Fig.~\ref{Fig1}(f)) transforms to $d_L(t)$.
Both the Gabor and Morlet transforms display separate broad lines regarding to the odd harmonics in the HHG process, consistent with the power spectrum in Fig.~\ref{Fig1}(c). Note that because of the adaptive frequency feature in wavelet transforms \cite{Ten_Lectures_on_Wavelets}, the frequency resolution for the harmonics below 11 is improved in Fig.~\ref{Fig1}(e), while the harmonics on the upper TF plane remain broaden.
The WV transform reveals signals between the odd harmonics, which is inconsistent with the features in the power spectrum and violates the parity symmetry in physics, due to the artificial interference inherited in the algorithm \cite{Book_TF_analysis}. Furthermore, the three TF representations show the same characteristics: the highest number of harmonics occurs around t$=30$T where the laser intensity reaches its maximum. In addition to odd harmonics, the Gabor and Morlet transforms show a broaden line near the 9th harmonic (H9) in the beginning few cycle (t$<15$T), where the laser intensity is not strong enough to induce the high-order harmonic process (see the red upward arrows in Fig.~\ref{Fig1}(i)). Note that such line does not appear in Fig.~\ref{Fig1}(f). The broaden line near H9 could be the resonant absorption from the 1s-2p transition. Although the three TF representations shed light on the underlying physics, the detail structure of the broaden line near H9 is obscure.

To resolve the aforementioned issue and reduce the artifact in the Gabor transform due to the window function, we adopt the SST to explore the IF of $d_L(t)$. Compared with the three transforms, the SST (Fig.~\ref{Fig1}(g)) demonstrates clear and distinct odd harmonics. In the enlargement of the SST result around H9, (Fig.~\ref{Fig1}(h)), an evident line is located at the H8.756 precisely, which corresponds to the energies for 1s-2p transition ($\frac{1}{2}(1-\frac{1}{2^2})/\omega_0=8.756$ in a.u.), manifested as a slightly shorter peak overlapping the H9 of the power spectrum in Fig.~\ref{Fig1}(c). That is, the substructures within H9 is clarified. Note that the intensity of the H8.756 is small because it arises from the near resonance absorption (not resonance absorption), as illustrated in Fig.~\ref{Fig1}(i). As the laser field increased ($\rm{t}<12\rm{T}$), we observe a shifting from H8.756 to H8.700 (Fig.~\ref{Fig1}(h)), which corresponds to the so-called AC Stark effect, where the energy levels of 2s, 2$\rm{p}_x$, 2$\rm{p}_y$, and 2$\rm{p}_z$ shift and split due to the breaking of the symmetry by the electric field. 
The blue, green, red, and cyan lines denote the energy difference of 1s-2s, 1s-2$\rm{p}_{x}$, 1s-2$\rm{p}_{z}$, and 1s-2$\rm{p}_{y}$, respectively, computed by Floquet method \cite{sic_Floquet,Hsu_Single_Molecule}, in the unit of $\omega_0$. 
Note that the energy difference of 1s-2$\rm{p}_{x}$ and 1s-2$\rm{p}_{y}$ are the same. 
However, we only observe a shifting from H8.756 to H8.700, the 1s-2$\rm{p}_z$ transition, because of $\langle\mathrm{1s}|z|\mathrm{2s}\rangle$, $\langle\mathrm{1s}|z|\rm{2p}_{x}\rangle$, and $\langle\mathrm{1s}|z|\rm{2p}_{y}\rangle = 0$ (selection rules). The SST not only addresses the intrinsic blurring in the other three TF methods but also reveals fundamental physical processes -- the AC Stark effect and selection rule. The mechanism is depicted in Fig.~\ref{Fig1}(i).

The jump from H8.700 to H9 in Fig.~\ref{Fig1}(g) results from the fact that the signals (H8.756) caused by the atomic structure (1s-2p near resonance absorption) was overtaken by the transition between the dressed states and present high-order harmonics. Note that in a strong field, the dynamics is dominated by the transition between dressed states $|M,N\rangle$ formed by the electron state $M$ and the photon state $N$, as shown in Fig.~\ref{Fig1}(i).

Similarly, the SST can indicate the line H10.38, despite weak, corresponding to the emission of the 1s-3p transition ($\frac{1}{2}(1-\frac{1}{3^2})/\omega_0=10.38$), in contrast with Fig.~\ref{Fig1}(d) and (e). Note that the line in Fig.~\ref{Fig1}(e) is more pronounced due to a coefficient proportional to $\omega$ in the Morlet transform \cite{Ten_Lectures_on_Wavelets, Tong_Morlet}. 


When the laser field comes to an end, Fig.~\ref{Fig1}(g) shows spectral lines located in H8.756, H10.377, and H12.07 harmonics. The reason of the existence of the H12.07 is not clear (it may be originated from the superposition of the excited states \cite{Antoine_time_profile_harmonics}), but this spectral lines can also be found the in the Gabor and Morlet transforms and the power spectrum.

To summarize, the synchrosqueezing transform is employed for the first time to analyze a quantum dynamical system, a hydrogen atom in a laser field based on an {\it ab inito} level simulation. The algorithm and theory of the SST have been briefly presented in this study. Moreover, compared with the Gabor, Morlet, and Wigner-Ville transforms, the SST reveals the accurate and detail features of the AC Stark effect and high harmonic generation. In particular, the signal shifting from 8.756 to 8.700 exactly corresponds to the energy difference of 1s-2$\textrm{p}_z$ transtion, consistent with the selection rules and Floquet computation. We believe that in addition to the high harmonic generation, the SST can be applied to explore other quantum dynamical processes, \textit{e.g.}, nuclear magnetic resonance. We also hope that this work will motivate additional studies to explore fundamental optical processes.

This work was supported by the National Science Council and National Taiwan University (Grant No. 103R8700-2 and 103R104021). The authors would also like to thank Dr. Tak-San Ho from Department of Chemistry in Princeton University and Dr. Han-Chih Chang from Department of Physics in University of Washington at Seattle for their valuable suggestions.

\pagebreak

\end{document}